# The multiobjective multidimensional knapsack problem: a survey and a new approach

T. Lust[1], J. Teghem

Université of Mons - Faculté Polytechnique de Mons
Laboratory of Mathematics and Operational Research
20, place du parc 7000 Mons (Belgium)

**Abstract:** The knapsack problem (KP) and its multidimensional version (MKP) are basic problems in combinatorial optimization. In this paper we consider their multiobjective extension (MOKP and MOMKP), for which the aim is to obtain or to approximate the set of efficient solutions. In a first step, we classify and describe briefly the existing works, that are essentially based on the use of metaheuristics. In a second step, we propose the adaptation of the two-phase Pareto local search (2PPLS) to the resolution of the MOMKP. With this aim, we use a very-large scale neighborhood (VLSN) in the second phase of the method, that is the Pareto local search. We compare our results to state-of-the-art results and we show that we obtain results never reached before by heuristics, for the biobjective instances. Finally we consider the extension to three-objective instances.

## 1 Introduction

Since the 70ies, *multiobjective optimization* problems (MOP) became an important field of operations research. In many real applications, there exists effectively more than one objective to be taken into account to evaluate the quality of the feasible solutions.

A MOP is defined as follows:

$$\text{``max''} \qquad z(x) = z_k(x) \quad k = 1, \ldots, p \quad \text{(MOP)}$$
$$\text{s.t} \qquad x \in X \subset \mathbb{R}^n_+$$

where $n$ is the number of variables, $z_k(x) : \mathbb{R}^n_+ \to \mathbb{R}$ represents the $k^{th}$ objective function and $X$ is the set of feasible solutions. We will denote by $Z = \{z(x) : x \in X\} \subset \mathbb{R}^p$ the image of $X$ in the objective space.

Due to the typically conflictive objectives, the notion of optimal solution does not exist generally anymore for MOPs. However, based on the dominance relation of Pareto (see Definition 1), the notion of optimal solution can be replaced by the notion of efficient (or Pareto optimal) solution (see Definition 2).

**Definition 1** *A vector $z \in Z$ dominates a vector $z' \in Z$ if, and only if, $z_k \leq z'_k$, $\forall\, k \in \{1, \ldots, p\}$, with at least one index $k$ for which the inequality is strict. We denote this dominance relation by $z \prec z'$.*

**Definition 2** *A feasible solution $x \in X$ is efficient if there does not exist any other solution $x' \in X$ such that $z(x') \prec z(x)$. The image of an efficient solution in objective space is called a non-dominated point.*

---

[1]Corresponding author. Address: thibaut.lust@umons.ac.be.

In the following, we will denote by $X_E$, called efficient set, the set of all efficient solutions and by $Z_N$, called the Pareto front, the image of $X_E$ in the objective space.

Even if other approaches exist to tackle a MOP problem (aggregation of the objectives with a utility function, hierarchy of the objectives, goal programming, interactive method to generate a " good compromise ": see [73]), in this paper we are only interested in the determination, or the approximation, of $X_E$ and $Z_N$. It should be noted that in all heuristics presented in this paper, only an approximation of a minimal complete set [38] is determined: no equivalent solution generated will thus be retained.

It is first the problems with continuous variables which called the attention of the researchers: see the book of Steuer [71] for multiobjective *linear programming* (MOLP) problems and of Miettinen [65] for multiobjective *non-linear programming* [65] (MONLP) problems.

However it is well-known that discrete variables are often unavoidable in the modeling of many applications and for such problems the determination of $X_E$ and $Z_N$ becomes more difficult.

Let us consider for instance a multiobjective *integer linear programming* (MOILP) problem of the form:

$$\text{``max''} \qquad z(x) = z_k(x) = c^k x \quad k = 1, \ldots, p \quad \text{(MOILP)}$$
$$\text{s.t} \qquad x \in X = \{x \in \mathbb{Z}_+^n : Ax = b\}$$

In such case, we can distinguish two types of efficient solutions:

- The supported efficient solutions are optimal solutions of the weighted single-objective problem

$$\max \quad \sum_{k=1}^{p} \lambda_k z_k(x)$$
$$\text{s.t} \qquad x \in X$$

  where $\lambda \in \mathbb{R}_+^p$ is a weight vector with all positive components $\lambda_k$, $k = 1, \ldots, p$. We denote by $X_{SE}$ and $Z_{SN}$ respectively the set of supported efficient solutions and the set of corresponding non-dominated points in $\mathbb{R}^p$. The points of $Z_{SN}$ are located on the frontier of the convex hull of $Z$.

- Contradictorily with a MOLP problem, $Z_{SN}$ is generally a proper subset of $Z_N$ due to the non-convex character of $Z$: there exist efficient solutions which are non-supported. We denote by $X_{NE} = X_E \backslash X_{SE}$ and $Z_{NN} = Z_N \backslash Z_{SN}$ respectively the set of non-supported efficient solutions and the set of the corresponding non-dominated points in $\mathbb{R}^p$.

Already in the 80ies, several methods have been proposed to generate $X_E$ for MOILP problems [74]. The two main difficulties to overcome are that:

- The sets $X_E$ and $Z_N$, formed of discrete points, can be of very large cardinality;

- The sets $X_{NE}$ and $Z_{NN}$ are more difficult to determine.

Later, various multiobjective *combinatorial* optimization (MOCO) problems have been considered. Most of them are of the type

$$\text{``max''} \qquad z(x) = c^k(x) \quad k = 1, \ldots, p \quad \text{(MOCO)}$$
$$\text{s.t} \qquad x \in X = D \cap \{0, 1\}^n$$

where $D$ is a specific polytope characterizing the particular CO problem.

During the last 15 years, there has been a notable increase of the number of studies on MOCO problems. From the first survey [77] in 1994 till [24] in 2002, a lot of papers have been published and this flow is still increasing. The main reason of this phenomenon is the success story of the metaheuristics [35].

Effectively, it is quite difficult to determine exactly the sets $X_E$ and $Z_N$ for MOCO problems. This is a $\mathcal{NP}$-Hard problem even for CO problems for which a polynomial algorithm for the single-objective version exists such as linear assignment problem. Therefore, there exist only few exact methods able to determine the sets $X_E$ and $Z_N$ and we can expect to apply these methods only for small instances. For this reason, many methods are heuristic methods which produce approximations $\widehat{X_E}$ and $\widehat{Z_N}$ to the sets $X_E$ and $Z_N$. Due to the succes of metaheuristics for single-objective CO, multiobjective *metaheuristics* (MOMHs) became quickly a classic tool to tackle MOCO problems and it is presently a real challenge for the researchers to improve the results previously obtained.

The two main difficulties of MOMHs are related to the basic needs of any metaheuristics [35]:

- To assure sufficient intensity, i.e. to produce an approximation $\widehat{Z_N}$ as close as possible to $Z_N$;

- To assure sufficient diversity, i.e. to cover with $\widehat{Z_N}$ all the parts of $Z_N$.

Unfortunately, measuring the quality of an approximation or to compare the approximations obtained by various methods remains a difficult task: the problem of the quality assessment of the results of a MOMH is in fact also a multicriteria problem. Consequently, several indicators have been introduced in the literature to measure the quality of an approximation (see [89] for instance).

Some of them are unary indicators:

- The hypervolume $\mathcal{H}$ (to be maximized) [86]: the volume of the dominated space defined by $\widehat{Z_N}$, limited by a reference point.

- The $R$ measure (to be minimized) [45]: evaluation of $\widehat{Z_N}$ by the expected value of the weighted Tchebycheff utility function over a set of normalized weight vectors.

- The average distance $D_1$ and maximal distance $D_2$ (to be minimized) [15, 80] between the points of a reference set and the points of $\widehat{Z_N}$, by using the Euclidean distance. Ideally, the reference set is $Z_N$ itself, but generally it is not available; otherwise, it can be the non-dominated points existing among the union of various sets $\widehat{Z_N}$ generated by several methods, or an upper bound of $Z_N$ [25].

- The $\epsilon$ factor $I_{\epsilon 1}$ (to be minimized) by which the approximation $A$ is worse than a reference set $B$ with respect to all objectives:

$$I_{\epsilon 1}(A, B) = \inf_{\epsilon \in \mathbb{R}^+} \{ \forall z \in B, \exists z' \in A : z_k' \cdot \epsilon \geq z_k, \, k = 1, \dots, p \}$$

- The proportion $P_{Y_N}$ (to be maximized) of non-dominated points generated.

Unfortunately, none of these indicators allows to conclude that an approximation is better than another one (see [91] for details). Nevertheless, an approximation that finds better values for these indicators is generally preferred to others.

In the first time, the MOCO problems treated in the literature are those for which the single-objective versions belong to the class $\mathcal{P}$, like linear assignment and shortest path problems. In a second time, a large attention has been devoted first to the multiobjective knapsack problem (MOKP) and after to its multidimensional version (MOMKP). In the next section we survey briefly the existing literature on these two problems. We then present the new heuristic approach (section 3), the data used (section 4) and the results obtained (section 5).

## 2 The multiobjective knapsack literature

The single-objective knapsack problem is certainly one of the most studied $\mathcal{NP}$-Hard CO problem. In the book of Martello and Toth [61] and the most recent of Kellerer *et al.* [48], various methods—essentially branch and bound and dynamic programming approaches—are analyzed, for the KP and MKP and for some of its variants; in [48], a chapter is devoted to approximation algorithms for the KP and another presents different heuristics for the MKP.

We recall the formulation of the multiobjective MKP (MOMKP): given $n$ items ($i = 1, \ldots, n$) having m characteristics $w_j^i$ ($j = 1, \ldots, m$)—like weight, volume, ...—and $p$ profits $c_k^i$ ($k = 1, \ldots, p$), some items should be selected to maximize the $p$ total profits while not exceeding the $m$ knapsack capacities $W_j$ regarding the different characteristics.

The MOMKP problem is formulated as follows:

$$\text{``max''} \quad z_k(x) = \sum_{i=1}^{n} c_k^i x_i \qquad k = 1, \ldots, p$$

$$\text{subject to} \quad \sum_{i=1}^{n} w_j^i x_i \leq W_j \qquad j = 1, \ldots, m$$

$$x_i \in \{0, 1\} \qquad i = 1, \ldots, n$$

where $x_i = 1$ means that the item $i$ is selected to be in the knapsack. It is assumed that all coefficients $c_k^i$, $w_j^i$ and $W_j$ are nonnegative.

The majority of the instances treated in the following cited papers concern the biobjective case ($p = 2$), sometimes the three-objective case ($p = 3$).

The particular case of a single constraint ($m = 1$) corresponds to the MOKP.

We first recall a simple and important result for the single-objective KP ($m = p = 1$), due to Dantzig [20]. If the items are ordered by non-increasing efficiencies $\frac{c^i}{w^i}$, an optimal solution $\hat{x}$ of the linear relaxation of the KP is given by

$$\widehat{x}_i = \begin{cases} 1 & \text{if } i < s \\ \frac{W - \sum_{j=1}^{s-1} w^j}{w^s} & \text{if } i = s \\ 0 & \text{if } i > s \end{cases}$$

where the "split" item $s$ is defined by:

$$\sum_{i=1}^{s-1} w^i \leq W < \sum_{i=1}^{s} w^i$$

In 1980, Balas and Zemel [7] noticed that the optimal solution for random instances of the KP was very close to the optimal solution of the linear relaxation of the KP. The notion of *core* has thus been defined and the more efficient algorithms for solving the single-objective KP are based on this concept.

We recall that a core $\mathcal{C}$ is a subset of items defined around the split item, that is

$$C = \{i \,|\, n_1 \leq i \leq n_2\} \text{ with } n_1 < s < n_2$$

Only the variables $x_i, i \in \mathcal{C}$ are considered, the others are fixed respectively to 1 if $i < n_1$ and to 0 if $i > n_2$.

### 2.1 The MOKP

We first survey the literature concerning the MOKP ($m = 1$). We decompose the presentation of the methods in three groups (sub-section 2.1.1 till 2.1.3): exact methods, approximation algorithms and heuristic methods based on metaheuristics. We end this section by reviewing particular studies (sub-section 2.1.4). Please note that is impossible in this survey to present the basic elements of the metaheuristics cited in this section; for a description of these metaheuristics, the reader can refer for instance to the books of Glover and Kochenberger [35] or the recent one of Talbi [72]. Obviously it is the same for the presentation of the numerous adaptations of metaheuristics to the multiobjective framework. Nevertheless we indicate in the bibliography the references of the corresponding papers.

### 2.1.1 Exact methods

We can distinguish four approaches.

a) The two-phase method

Such method, proposed by Ulungu and Teghem [78], is initially dedicated to solve biobjective CO problems. The first phase generates the set $X_{SE}$ of supported efficient solutions by solving a sequence of single-objective KP obtained by several linear aggregations of the two objective functions. The second phase explores the interior of all the right triangles defined by two consecutive points of $X_{SE}$ to obtain the set $X_{NE}$ of non-supported efficient solutions. This second phase is more complex and is problem dependent. In 1997, Ulungu and Teghem [79] applied this method to a biobjective MOKP using a branch and bound method in the second phase. This method is improved by Visée *et al.* [83] in 1998 and these authors were able to solve randomly generated instances with up to 500 items. Jorge and Gandibleux [47] proposed in 2007 an improvement by using better bounds in the branch and bound and a ranking algorithm in the second phase.

b) Transformation into biobjective shortest path problems

This approach was proposed in 2003 by Captivo *et al.* [14]. The obtained biobjective shortest path problem was solved with an adapted version of the labeling algorithm of Martins *et al.* [62]. They used biobjective instances of three different types: random, weakly correlated and strongly correlated objectives. They solved these three types of instances with respectively up to 320 items, 800 and 900 items. They showed that they obtain better results (in term of computational time) than the method of Visée *et al.*. This approach was further extended by Figuera *et al.* [27].

c) Dynamic programming

A first attempt to extend a dynamic programming (DP) algorithm to the MOKP was suggested in 2000 by Klamroth and Wiecek [49]. Recently, in 2009, Bazgan *et al.* [11] developed this idea adding several complementary dominance relations to discard partial solutions. They obtained an efficient method, tested on different types of instances: random (type A), strongly correlated (type B), weakly uncorrelated (type C) and strongly uncorrelated (type D) biobjective instances. In less than two hours of computational time, they solved biobjective instances of type A, B, C and D with respectively 700, 4000, 500 and 250 items. They compared their results with the method of Captivo *et al.* and with an $\epsilon$-constraint method [37] coupled with the ILOG Cplex 9.0 solver, and they obtained better results and solved instances of higher size. They also tested their method on three-objective instances. Due to the explosion of the cardinality of $X_E$, they could only solve instances of type A with up to 110 items and of type C with up to 60 items. It is still remarkable since this is the first exact method that was adapted to three-objective instances of the MOKP.

d) Hybridization

In 2010, Delort and Spanjaard [23] proposed a hybridization (called *two-phasification*) between the two-phase method and a DP procedure for solving biobjective instances. The DP procedure is applied in the second phase. They also integrated shaving procedures and bound sets to reduce the size of the problems. They obtain better results than Bazgan *et al.* on random and correlated instances, and comparable results on uncorrelated instances.

### 2.1.2 Approximation algorithms

Following [26, 81], we first recall some definitions (in case of maximization).

- An *$\epsilon$-efficient set* is a set $Y_\epsilon$ of feasible solutions such that for all $x_e \in X_E$ there exists $y \in Y_\epsilon$ satisfying
$$f_k(y)(1+\epsilon) \geq f_k(x_e) \quad \forall k \in \{1, \ldots, p\}$$
with the same ratio $r = (1+\epsilon)$ for all objectives.

- A *$(1+\epsilon)$-approximation algorithm $A_\epsilon$* is an algorithm producing an $\epsilon$-efficient set in polynomial time.

- A *polynomial-time approximation scheme* (PTAS) is a family of algorithms that contains, for each $\epsilon > 0$, a $(1+\epsilon)$-approximation algorithm $A_\epsilon$.

- A *fully polynomial-time approximation scheme* (FPTAS) is a PTAS for which $A_\epsilon$ is polynomial in $\epsilon^{-1}$.

Few papers deal with approximation algorithms:

- In 2002, Erlebach *et al.* [26] presented an efficient and applicable FPTAS for the MOKP. They also presented a PTAS for the MOMKP based on linear programming.

- In 2006, Kumar and Banerjee [53] proposed a restricted evolutionary multiobjective optimizer, called REMO, which gives $(1+\epsilon)$ approximations. It is based on a restricted mating pool with a separate archive to store the remaining population. They presented a rigorous running time analysis of the algorithm.

- Recently, in 2009, Bazgan *et al.* [10] proposed a new FPTAS with a practical behavior. The main idea is the same that in their exact method [11]. They compared their FPTAS with the one of Erlebach *et al.* and they showed that they obtain better results.

### 2.1.3 Heuristic methods

Here the aim is to find a good approximation of $X_E$, generally using metaheuristics.

a) Simulated Annealing (SA)

- In 1993, Ulungu [76] presented in his thesis the first adaptation of SA to MO through MOSA [80]. SA is simply applied several times with a well-diversified set of weight sets aggregating the objective functions. For each weight set, the non-dominated solutions are kept and all these solutions are finally merged to achieve a unique approximation. He solved random biobjective instances with up to 500 items (the same instances than in the exact method of Visée *et al.*)

- In 1998, Czyzak and Jaszkiewicz [15] proposed the Pareto simulated annealing (PSA) to solve random biobjective, three and four objective instances with up to 800 items. In PSA, an initial set of solutions is generated. Weight sets are associated to each of these solutions that are optimized in the same way as in MOSA. For a given solution, the weight set is changed in order to induce a repulsion mechanism assuring dispersion of the solutions over all the regions of the Pareto front.

b) Tabu Search (TS)

- Ben Abdelaziz and Krichen [13] proposed in 1997 a tabu search-based method. They solved random biobjective instances with up to 100 items. They extended their method in 1999 by integrating the tabu search into a genetic algorithm [1].

- Gandibleux and Freville [29] proposed in 2000 a TS, called MOTS. In MOTS, an augmented weighted Tchebycheff scalarizing function is used to select the best neighbor. The weight set is dynamically updated such that the search process is oriented towards a region of the objective space where few non-dominated points have been generated. The authors also used different techniques to reduce the decision space to an interesting area. They tested MOTS on random biobjective instances of 50 and 100 items.

c) Genetic Algorithm (GA)

Gandibleux *et al.* [31] in 2001 developed a two-phase method: in the first phase an exact algorithm to compute the set $X_{SE}$ has been used and in the second phase the traditional components of a GA has been applied.

d) Scatter Search (SS)

- Gomes da Silva *et al.* proposed in 2006 [17] a scatter search based method, following the usual structure of SS. They tested their method on large size random biobjective instances with a number of items going from 100 to 6000. They compared their method to the exact method of Visée *et al.* and showed that for the 100, 300, 500 items instances, they generate in average respectively 33.13%, 9.75% and 5.12% of the efficent solutions. Obvioulsy, the running time of their method is much lower: for $n = 500$, the exact method takes about 400 times the computational time required by their heuristic SS.

- In 2007 [19], the same authors presented an improvement of their previous method, by modifying some elements of the SS technique. In particular in the solution combination method which combines solutions from each subset of a reference set to create new solutions, they used an adapted version of the exact method of Visée *et al.* to solve small size residual problems. They improved their previous results since they generate, for the same instances with 100, 300 and 500 items, respectively 87.3%, 95.0% and 91.2% of the efficient solutions. On the other hand, the improvement of running time was not anymore spectacular and for the 500 items instance, the ratio of running times is only equal to 1.5 instead of 400. They also showed that they obtain better results that those generated by the MOSA of Ulungu *et al.* and the genetic approach of Gandibleux *et al.* presented above.

e) Linear Relaxation

Zhang and Ong [85] proposed in 2004 a simple method essentially based on an efficient heuristic for the linear relaxation. They tested their method on random biobjective instances. They solved instances from 500 to 50000 items and they showed that they obtain better results than if their method is coupled with the ILOG CPLEX 7.5 solver.

### 2.1.4 Particular studies

Other studies concern the MOKP.

- In 2006, Gandibleux and Klamroth [30] studied cardinality bounds for the MOKP based on weighted sum scalarizations. They showed that we can use these bounds to reduce the feasible set of the biobjective MOKP.

- In 2007, Gandibleux and Ehrgott [25] introduced the concept of bound sets for MOCO problems. Indeed, well-known bounds for multiobjective problems are the ideal point (lower bound) and the nadir point (upper bound) but first of all, these bounds are not easy to compute, especially the nadir point, and secondly, these values are very far from the Pareto front. They thus generalized the notion of a bound value to that of a bound set and applied these concepts to, among others, the MOKP. They obtain upper bound set by using the linear relaxation of the MOKP and lower bound set by using a simple greedy algorithm.

- In 2007, Jorge *et al.* [47] presented new properties aiming to a priori reduce the size of the biobjective MOKP. Based on a lower and upper bound on the cardinality of a feasible solution for KP introduced by Glover in 1965 [32] and on dominance relations in the space of data of the MOKP, they reduced the size of the biobjective instances of the MOKP by fixing, a priori, about 10% of the variables, on random instances.

- Gomes da Silva *et al.* recently studied, in 2008, the interesting notion of core problems for the MOKP [18]. Indeed, this concept has never been directly used in the MO framework. They thus investigated the existence of the core structure in the MOKP and defined the notion of *biobjective* core. They reported computational experiments related to the size of the biobjective core on different types of instances. The results show that, on average, the biobjective core is a very small percentage of the total number of items. Then, they proposed a heuristic and an exact method based on these results, but without experimenting these methods.

## 2.2 The MOMKP

For the MOMKP, as far as we know, this is mainly heuristic methods that have been developed. Despite many of them are hybrid methods, we try to make a distinction between those mainly based on evolutionary algorithms (sub-section 2.2.1) and those mainly based on local search (sub-section 2.2.2). Some particular studies are analyzed in sub-section 2.2.3.

### 2.2.1 Evolutionary algorithms

- It seems that is first Zitzler and Thiele [90] who tackled the MOMKP, in 1999. They performed a comparative study of five different MOGAs for the MOMKP, and also introduced SPEA. They showed that this method outperforms the other methods. In this paper, they introduced the instances of the MOMKP that will be used by many other authors later. For these instances (that we will call the ZMKP instances), the number of objectives is equal to the number of constraints ($m = p$). Nine different instances with two, three and four objectives, in combination with 250, 500 and 750 items have been created. The profits and the weights were randomly generated in the interval [10,100]. The knapsack capacities were set to half the total corresponding weight. In all the works presented below, if nothing is mentioned for the instances used, that means that the ZMKP instances are considered. An improved version of the SPEA algorithm, called SPEA2, was further developed by Zitzler *et al.* in 2001 [88]. In SPEA2, the rank of an individual takes into account the number of individuals that dominate it and the number of individuals dominated by the individual. Clustering in the objective space is used in order to reduce the number of individuals and allows to obtain only one representative individual in small regions of the objective space. The new algorithm has been tested on a subset of the ZMKP instances and the authors concluded that SPEA2 performs better than its predecessor on all instances.

- In 2000, Knowles and Corne [51] compared M-PAES [52], based on Pareto ranking of the solutions with RD-MOGLS [40], based on random scalarization functions. They showed that both algorithms work well and produce better results than (1+1)-PAES [50].

- In 2000, Jaszkiewicz [42] applied MOGLS and compared it to SPEA. In MOGLS, the best solutions in the population $P$ for the scalarizing function form a temporary population $TP$ of small size. Both parents are randomly select in $TP$. He showed that MOGLS outperforms SPEA. In 2001, Jaszkiewicz continued his experiments and in [44], he compared five algorithms: MOGLS, M-PAES [51], SMOSA [70], MOSA [80] and PSA [15]. Jaszkiewicz concluded that MOGLS outperforms the other methods. In [45], a comparison has been realized between MOGLS, SPEA, M-PAES and IMMOGLS [39] and he also concluded that MOGLS outperforms the other methods. Jaszkiewicz published the results obtained with MOGLS and these results became the new reference for testing new methods.

- In 2004, Jaszkiewicz [46] gave up his MOGLS algorithm and compared three MOMHs: his PMA [43], SPEA and the controlled elitist non-dominated sorting genetic Algorithm (CENSGA) [21]. In PMA, the selection is based on a tournament. The two parents selected are the winners of a tournament between solutions coming from a sample of size $T$ randomly drawn from the population. The size of $T$ is set in order to guarantee that this selection procedure gives the same quality of offsprings than with the selection procedure of MOGLS. Indeed, PMA has been developed by Jaszkiewicz in order to reduce the running time of the selection process of its MOGLS algorithm, while keeping same quality results. PMA has been adapted to the MOMKP in a more sophisticated way than MOGLS. SPEA and CENSGA have also been adapted to the MOMKP by Jaszkiewicz in the same way as PMA. The three methods share thus some identical components. Jaszkiewicz showed that PMA performs better than SPEA and CENSGA on instances with more than two objectives. After this work, it seems that this adaptation of PMA became the new reference method, since this method is an evolution of MOGLS and is adapted with more sophisticated operators. Unfortunately, the results of PMA were not published, and the previous results obtained with MOGLS remained the reference, even if it was possible to generate the results of PMA with the source code of PMA that Jaszkiewicz published, through the MOMHLib++ library [41]. This library makes it possible to run various existing MOMHs on the MOMKP. In this library, MOGLS has also been adapted following PMA and gives better results than the initial MOGLS [42].

- Alves and Almeida [5] presented in 2007 a genetic algorithm based on the Tchebycheff scalarizing function, called multiobjective Tchebycheff genetic algorithm (MOTGA). Several stages were performed; each one intended for generating potentially non-dominated points in different parts of the Pareto front. Different weight vectors act as pivots to define the weighted Tchebycheff scalarizing functions and direct the search for each stage. Linear relaxation of the integer program based on weighted Tchebycheff scalarizing functions followed by a repair procedure has been used to produce initial solutions. They compared MOTGA to SPEA, SPEA2 and MOGLS and they showed that MOTGA obtains better quality indicators than the other MOMHs, with a lower running time, even if the number of potentially non-dominated solutions obtained by MOTGA is not very high.

- In 2008, Lust and Teghem [59] presented a memetic algorithm integrating tabu search (MEMOTS). The particularity of the method is in the selection of the parents for recombination. They used a dynamic hypergrid created in the objective space to select parents located in a region of minimal density. Then they apply a tabu search to the offspring. They showed that they obtain better results than MOGLS and PMA for various indicators for the instances with two or three objectives.

### 2.2.2 Local Search

- In 2002, Barichard and Hao [9] proposed a TS that gives better results than SPEA but worse results than MOGLS. Their method is a classic TS with a Tchebycheff scalarizing function to select the best neighbor, except that an interesting way to measure the diversity has been added. They improved in 2003 the TS by integrating it into a GA [8] and with this new hybrid method, they concluded that they obtain better results than MOGLS.

- Li *et al.* [56] studied in 2004 a hybrid adaptation of the estimation of distribution algorithm (EDA) [54, 67] to the MOMKP, by using a local search based on weighted sums, a random repair method and a population sampled from a probability distribution. They compared their results with those of MOGLS and they showed that they obtain better results for some indicators.

- In 2004, Vianna and Arroyo [82] proposed an adaptation of the GRASP metaheuristic to the MOMKP. The adaptation of GRASP follows this frame: at each iteration a weighted linear sum of the objectives is defined and a solution is built considering this linear sum. The solution is then improved by a local search that also makes use of the linear sum. They compared their method to SPEA2 and MOGLS and showed that they obtain better results.

- Gomes da Silva *et al.* [16] adapted their SS for the MOKP [17] to also tackle the MOMKP. Surrogate relaxation was used to convert the MOMKP into a MOKP by generating adequate surrogate multipliers. They first compared their method with SPEA and MOGLS on the ZMKP instances with only two objectives. They showed by comparing in objective space the potentially non-dominated points generated that they obtain better results on these instances. They then tested their method on new instances, of bigger size: the number of items was included between 1000 and 3000 and the number of constraints went from 10 to 100. The values of the profits and the weights were randomly generated between 1 and 100 and each knapsack constraint capacity was equal to 50% of the sum of their weights. They evaluated the quality of their results by measuring different distances between the approximations obtained and the upper bound obtained with the surrogate relaxation. They showed that the approximations generated are very close to the upper bound. On the other hand, the running time was close to one hour, but given the size and the number of constraints taken into account, that remains reasonable.

- In 2009, Alsheddy and Tsang [3] proposed a guided Pareto local search (GPLS). This method combines GLS (guided LS) of Voudouris and Tsang [84] with a penalization of the objective functions to escape from local optima, and PLS of Paquete *et al.* [66]. They compared their method with general MOEA (SPEA2 and NSGA-II [22]). In [4], they improved their approach by using better initial solutions.

### 2.2.3 Particular Studies

We end the review of the works concerning the MOMKP by four particular studies.

- In 2007, Sato *et al.* [69] studied and compared the effects on performance of local dominance and local recombination applied with different locality. They used the NSGA-II algorithm [22] to run their experiments. They used the MOMKP to compare the different approaches and they showed that the optimum locality of dominance is different from the optimum locality of recombination. They also pointed out that the performances when local dominance and local recombination with different locality are applied, are significantly better than the performances when local dominance or local recombination is applied alone, or even when local recombination and dominance with the same locality are applied.

- In 2008, Beausoleil *et al.* [12] applied a multi-start search method combined with path-relinking to generate potentially efficient solutions in a so-called *balanced* zone of the Pareto front. However, this concept is not defined at all: it seems that it means that the method focuses on solutions that establish a good compromise between all the objectives. The method is relatively sophisticated and integrates many ideas developed by Glover [33, 34]: ghost image process, strategic oscillation approach and use of conditional-exclusion memory or developed by Glover and Laguna [36] with the path-relinking technique. In addition, the method presents many parameters. The results of the method are compared to many MOGAs, including SPEA, SPEA2, NSGA and NSGA-II but not with MOGLS or PMA which gives better performances on the MOMKP. They showed that their approach is competitive regarding those algorithms.

- Florios *et al.* [28] published in 2010 an exact method for solving the MOMKP. The method is a simple branch and bound method based on the ideal point as fathom rule and on branching heuristics. They experimented their method on instances with $p = m = 3$. They have compared the results with the "$\epsilon$-constraint" approach of Laumanns *et al.* [55] and they show that their approach is faster.

- In 2009, Mavrotas *et al.* [63] adapted the notion of core to the MOMKP, as Gomes da Silva *et al.* did for the MOKP. They used linear relaxations to define different weight intervals that allow to create different MOMKP problems, of smaller size than the original. For that, they used the core concept developed by Puchinger *et al* for solving the MKP [68]. They used the exact method developed in [28] to solve the restricted problems. They used biobjective instances, with $n$ going from 6 to 50, and $m$ from 5 to 10. They also used some of the ZMKP instances. They studied the influence of the size of the core to the quality of the results and the computational time. For example, for the ZMKP instance of 250 items, with 2 constraints and 2 objectives, they can generate 81% of $X_E$ in about 30 minutes. But they need 21 hours to generate 93% of $X_E$.

## 3 New approach

### 3.1 Presentation

From the preceding survey, the most promising methods are those based on the notion of core. The adaptations of Mavrotas *et al.* [63] of the method proposed by Gomes da Silva *et al.* [18] allow to obtain high quality results, but unfortunately with a substantial computational time. This high computational time comes essentially from the trend to try to computing exactly the core or to solving exactly the residual problems. We show with this new approach, that is possible to use these notions, in a complete heuristic way, that allows to obtain in the same time: high quality results and small computational times.

We have used the two-phase Pareto local search (2PPLS) [60], coupled with a very-large scale neighborhood (VLSN) [2]. With 2PPLS, as with methods based on the notion of core, the "divide and conquer" principle is applied by solving many residual problems. However, all those residual problems will be managed by the VLSN and selected in a heuristic way.

Local search with VLSN uses a large neighborhood combined with an efficient method to explore the neighborhood (otherwise it takes too much time to explore the neighborhood). This technique is very popular in single-objective optimization. For example, one of the best heuristics for solving the single-objective traveling salesman problem (TSP), the Lin-Kernighan heuristic [57], is based on VLSN. On the other hand, there is almost no study of VLSN for solving MOCO problems. The only known result is the local search of Angel *et al.* [6], which integrates a dynasearch neighborhood (the neighborhood is solved with dynamic programming) to solve the biobjective TSP.

We now present 2PPLS. This method only needs an initial population and a neighborhood (the VLSN in our case, that will be present later).

The aim of the first phase of 2PPLS is to generate a good approximation of the supported efficient solutions, by using weighted sums and efficient single-objective solvers for optimizing the weighted single-objective problems. The aim of the second phase is to generate non-supported efficient solutions. For that, we use the Pareto local search (PLS) [6, 66]. This method is a straightforward adaptation of local search to MO and only needs a neighborhood function $\mathcal{N}(x)$. At the end, a local optimum, defined in a MO context, is found [66] (called a Pareto local optimum set). The main part of the adaptation of 2PPLS to MOCO problems concerns the definition of the neighborhood (a VLSN in our case).

The pseudo-code of 2PPLS is given by Algorithm 1.

---

**Algorithm 1** 2PPLS

---

    Parameters ↓: an initial population $P_0$, a neighborhood function $\mathcal{N}(x)$.
    Parameters ↑: an approximation $\widetilde{X}_E$ of the efficient set $X_E$.

    --| Initialization of $\widetilde{X}_E$ and a population $P$ with the initial population $P_0$
    $\widetilde{X}_E \leftarrow P_0$
    $P \leftarrow P_0$
    --| Initialization of an auxiliary population $P_a$
    $P_a \leftarrow \varnothing$
    **while** $P \neq \varnothing$ **do**
        --| Generation of all neighbors $p'$ of each solution $p \in P$
        **for all** $p \in P$ **do**
            **for all** $p' \in \mathcal{N}(p)$ **do**
                **if** $f(p) \not\preceq f(p')$ **then**
                    `AddSolution`$(\widetilde{X}_E \updownarrow, p' \downarrow, f(p') \downarrow, Added \uparrow)$
                    **if** $Added = true$ **then**
                        `AddSolution`$(P_a \updownarrow, p' \downarrow, f(p') \downarrow)$
        --| $P$ is composed of the new potentially efficient solutions
        $P \leftarrow P_a$
        --| Reinitialization of $P_a$
        $P_a \leftarrow \varnothing$

---

The method starts with the population $P$ composed of potentially efficient solutions given by the initial population $P_0$. Then, all the neighbors $p'$ of each solution $p$ of $P$ are generated. If a neighbor $p'$ is not weakly dominated by the current solution $p$, we try to add the solution $p'$ to the approximation $\widetilde{X}_E$ of the efficient set, which is updated with the procedure `AddSolution`. This procedure is not described in this paper but simply consists of updating an approximation $\widetilde{X}_E$ of the efficient set when a new solution $p'$ is added to $\widetilde{X}_E$. This procedure has four parameters: the set $\widetilde{X}_E$ to actualize, the new solution $p'$, its evaluation $f(p')$ and a boolean variable called $Added$ that returns $True$ if the new solution has been added and $False$ otherwise. If the solution $p'$ has been added to $\widetilde{X}_E$, the boolean variable $Added$ is true and the solution $p'$ is added to an auxiliary population $P_a$, which is also updated with the procedure `AddSolution`. Therefore, $P_a$ is only composed of (new) potentially efficient solutions. Once all the neighbors of each solution of $P$ have been generated, the algorithm starts again, with $P$ equal to $P_a$, until $P = P_a = \varnothing$. The auxiliary population $P_a$ is used such that the neighborhood of each solution of the population $P$ is explored, even if some solutions of $P$ become dominated following the addition of a new solution to $P_a$. Thus, sometimes, neighbors are generated from a dominated solution.

## 3.2 Adaptation of 2PPLS to the MOMKP

The two-phase Pareto local search (2PPLS) only requires two elements to be adapted to a MOCO problem: an initial population and a neighborhood.

### 3.2.1 Initial population

We use a greedy heuristic. To create a new solution, the items are added to the knapsack one by one. At each iteration, the item $s$ that maximizes the following ratio ($R_1$) [64] is selected:

$$R_1 = \frac{\sum_{k=1}^{p} \lambda_k c_k^s}{\sum_{j=1}^{m} \left( \frac{w_j^s}{W_j - \sum_{i=1}^{n} w_j^i x_i + 1} \right)} \tag{1}$$

The greedy procedure is run $S$ times, $S$ being the number of weight sets $\lambda$ used. The weight sets are uniformly distributed for biobjective instances and randomly generated for three-objective instances.

### 3.2.2 Very-large scale neighborhood

The aim of the VLSN is to define the function $\mathcal{N}(x)$, that gives a set of neighbors from a current solution $x$.

For that, two lists are created, both of size $L$: one list (L1) containing the items candidates to be removed (thus present in $x$) and another list (L2) containing the items candidates to be added (thus missing in $x$).

To create L1, the items, in $x$, minimizing the ratio $R_2$, defined by

$$R_2 = \frac{\sum_{k=1}^{p} \lambda_k c_k^s}{\sum_{j=1}^{m} w_j^s} \tag{2}$$

are selected.

To create L2, the items, not in $x$, maximizing the ratio $R_1$ (defined by (1)) are selected. For biobjective instances, the weight set $\lambda$ necessary to the computation of these ratios is determined according to the relative performance of the potentially efficient solution $x$ selected, for the different objectives, among the population $P$. That is better the evaluation of the solution $x$ according to an objective is, higher is the value of the weight according to this objective. For three-objective instances, the weight set is randomly generated.

Once both lists containing each $L$ items have been created, we merge them to create a new MOMKP instance, called the residual problem, composed of $(L * 2)$ items. The capacities $W_j$ of the residual problem are equal to $W_j - \sum_{\substack{i=1 \\ i \notin L1}}^{n} w_j^i x_i$ with $j = 1, \ldots, m$.

We have then to solve the residual problem. As this problem is of small size, we can use an exact method. We have implemented a branch and bound method based on the method of Florios *et al.* [28]. In addition, to be able to use a higher value of $L$ while keeping reasonable running times, we have also used a heuristic method. For that, we have employed a simplified version of MEMOTS [59]: no hypergrid will be used to select the parents. The reason is that the number of potentially efficient solutions generated will not be high, and thus managing a hypergrid to select a solution of minimal density is not worthwhile. The advantage of not using the hypergrid is the simplification of the method and the elimination of parameters to tune. Consequently, both parents will be selected randomly in the set of potentially efficient solutions.

Once the residual problem has been solved, we merge the efficient solutions (or potentially efficient depending on the method used) of this small problem with the current solution, to obtain the neighbors.

# 4 Data and reference sets

As many authors previously did, we use the ZMKP instances with 250, 500 or 750 items, two objectives and two constraints or three objectives and three constraints. It should be noted that, thereafter, when we will speak about, for example, the 250-2 instance, it will mean the instance with 250 items and two objectives.

In order to assess the quality of the approximations generated, we have used the unary indicators presented in the introduction.

Some of these indicators need a reference set. For the biobjective instances, we use the non-dominated points generated by Tuyttens [75] by applying the $\epsilon$-constraint method coupled with the commercial CPLEX solver. For the three-objective instances, we approximate the efficient solutions by applying several times heuristic methods (the MEMOTS [59] and MOGLS [45] methods) during a high number of iterations. We then only retain the potentially non-dominated points.

The values of the reference points to compute the $R$ indicator are the same than the ones than Jaszkiewicz used in [42]. As also done by Jaszkiewicz in [42], the number of weight sets used to compute $R$ is equal to 201 for the biobjective instances and to 50 for the three-objective instances.

The bounding point considered to compute the hypervolume is simply the point (0,0), for the biobjective instances. For the three-objective instances, we did not compute the hypervolume since its computation was very time-consuming given the high size of the sets $\widehat{Z_N}$.

# 5 Results

The computer used for the experiments is a Pentium IV with 3 Ghz CPUs and 512 MB of RAM. Twenty runs of the algorithms are performed each time. The running time of our implementation of the algorithms corresponds to the wall clock time.

## 5.1 Study of the influence of the quality of the initial population

We have tried different values of $S$ (see sub-section 3.2.1) for the generation of the initial population. However, given the quality of the neighborhood used in the second phase, the value of $S$ had only a small influence on the quality of the results. We do not report the results here but you can find the results of this study in [58]. In the following, we will use $S = 100$ for the biobjective instances and $S = 150$ for the three-objective instances.

## 5.2 Study of the influence of the length of the neighborhood

It is interesting to measure the improvements of quality when the size of the neighborhood is increased.

We show in Figure 1 the evolution of the proportion of non-dominated points generated $P_{Y_N}$ and the running time (only of the second phase) according to $L$, when the branch and bound method is used to solve the residual problems (called the EXACT subroutine), for the 500-2 instance. We vary the value of $L$ from 4 to 8. We see that there are strong improvements of $P_{Y_N}$ when $L$ is increased. On the other hand, the running time evolves exponentially according to $L$. Using a value of $L$ superior to 8 would give unreasonable running times.

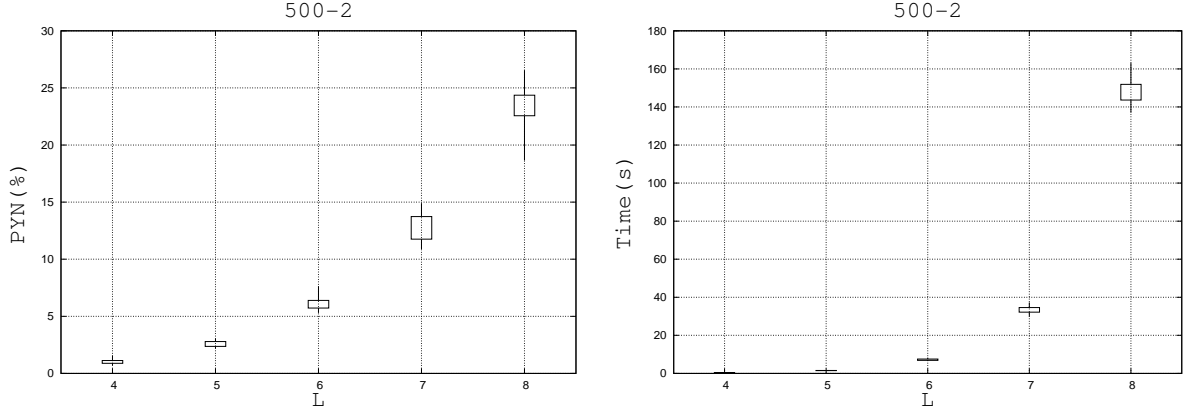

Figure 1: Influence of $L$ with the EXACT subroutine.

In Figure 2, we show the evolution of $P_{Y_N}$ and the running time according to $L$, in case of the use of MEMOTS to solve the residual problems, for the 500-2 instance. We vary the values of $L$ from 4 to 20. We use three different numbers of iterations for MEMOTS: $N = 100$, $N = 200$ and $N = 400$. We see that for small values of $L$, $P_{Y_N}$ is more or less equal no matter the number of iterations. From $L$ equal to 10, it is clear that we obtain better results if $N$ is higher. On the other hand, the running time is bigger when $N$ is higher, but still evolves more or less linearly according to $L$. An interesting behavior is pointed out by the figure showing the evolution of $P_{Y_N}$ according to $L$. From $L$ equal to about 16 and for a number of iterations $N$ equal to 100 or 200, there is a deterioration of $P_{Y_N}$ while the running time is increasing. It means that the number of iterations performed in MEMOTS is not high enough to solve the residual problems, and that therefore the quality of the approximations obtained for the residual problems is not good enough to improve $P_{Y_N}$. Fixing good values for $L$ and $N$ seems thus not easy since these two values have to be increased at the same time if we want to improve the qualities of the results.

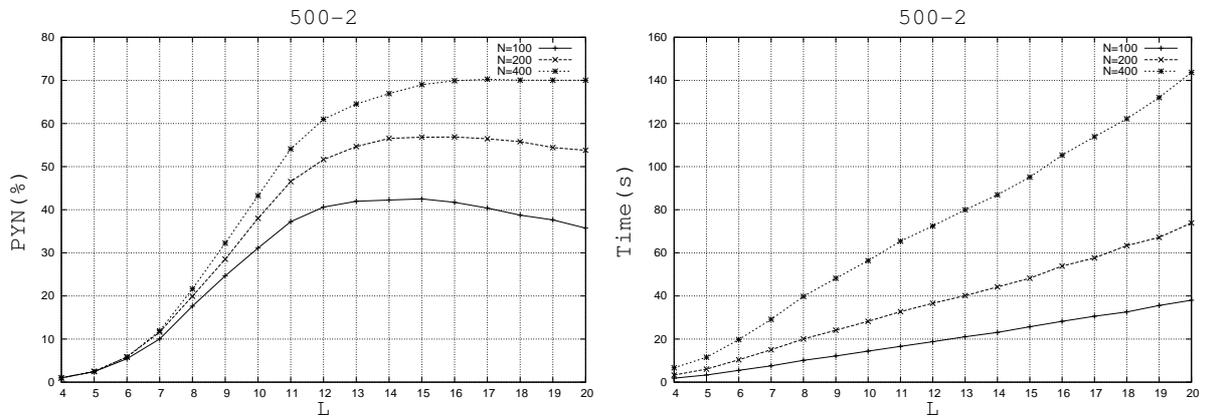

Figure 2: Influence of $L$ with the MEMOTS subroutine.

Now, it is interesting to see, for a fixed running time, what is the best to do: using the EXACT or MEMOTS subroutine, with which value of $L$, and for MEMOTS, with how many iterations.

In order to answer this question, we have represented in Figure 3, the evolution of $P_{Y_N}$ according to the running time (the running time is controlled by the value of $L$: from 4 to 8 for the EXACT subroutine and from 4 to 20 for the MEMOTS subroutine), for the 250-2 and 750-2 instances. We see that for only very small running times, it is better to use the EXACT subroutine. As soon as we give more running times, it is better to use the MEMOTS subroutine. In this case, the good combination between $L$ and $N$ has to be determined according to the running time given. The higher the running time allowed, the higher the number of iterations should be.

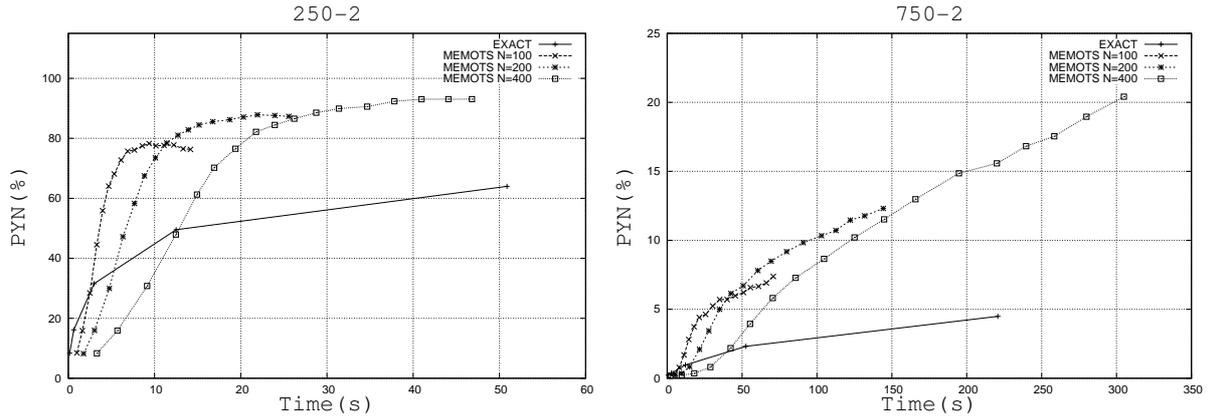

Figure 3: Comparison between the EXACT and MEMOTS subroutines to solve the residual problems according to the running time.

## 5.3 Comparison with other algorithms on biobjective instances

We have obtained the following results for the 250-2, 500-2, 750-2 instances: SPEA [90] (30 runs); SPEA2 [88] (30 runs, but only for the 750-2 instance); MOGLS00 [42] (20 runs); MOGLS04 [42] (20 runs, different than MOGLS00 since obtained with the library MOMHLib++ [41]); PMA [46] (20 runs); IMMOGLS [39] (20 runs); MOGTS [9] (1 run); GRASP [82] (1 run); MOTGA [5] (20 runs); PATH-RELINKING [12] (30 runs); GPLS [3] (30 runs); mGPLS [4] (30 runs); iGPLS [4] (30 runs). These results have been obtained either by downloading them from web sites or by asking them personally to the different authors. We see that we have obtained quite a lot of results. It is only a pity that we did not obtain the results of Gomes da Silva *et al.* [16] with their scatter search method.

Thank to these results, we have generated a reference set, called ALL, formed by merging the potentially non-dominated points obtained by all the runs of all algorithms, which gives a high quality set.

However, we show that is possible to obtain better results than this set, for all the indicators considered, in reasonable times, with 2PPLS and the VLSN solved with the MEMOTS subroutine. We have carefully selected the parameters such that we obtain better or equal results than the reference set ALL for all indicators. The parameters are the following:

- 250-2: $L = 9$ and $N = 200$.

- 500-2: $L = 15$ and $N = 100$.

- 750-2: $L = 9$ and $N = 100$.

The results for 2PPLS are given in Table 1. $|PE|$ gives the number of potentially non-dominated points generated. We see that we obtain better or equal values for all indicators, in very reasonable running times: 7s for 250-2, 23s for 500-2 and 18s for 750-2. 2PPLS with this configuration seems thus very competitive.

Table 1: Comparison between 2PPLS and ALL based on the indicators.

| Instance | Algorithm | $\mathcal{H}(10^7)$ | $I_{\epsilon 1}$ | $R$ | $D_1$ | $D_2$ | $|PE|$ | $P_{Y_N}(\%)$ | Time(s) |
|----------|-----------|---------------------|------------------|-----|-------|-------|--------|---------------|---------|
| 250-2    | 2PPLS     | **9.8690**          | **1.000508**     | **245.740328** | **0.029** | **2.680** | 482.10  | **68.05** | 7.27  |
|          | ALL       | **9.8690**          | 1.000839         | 246.129567     | 0.069     | 2.838     | 376.00  | 31.87     | /     |
| 500-2    | 2PPLS     | **40.7873**         | **1.000282**     | **430.902857** | **0.025** | **1.976** | 1131.00 | **42.85** | 23.43 |
|          | ALL       | 40.7850             | 1.000513         | 431.961766     | 0.081     | 2.045     | 688.00  | 5.51      | /     |
| 750-2    | 2PPLS     | **89.3485**         | **1.000509**     | **743.615748** | **0.076** | **1.494** | 1558.90 | **4.15**  | 17.52 |
|          | ALL       | 89.3449             | 1.000553         | 744.089577     | 0.092     | **1.494** | 996.00  | 0.99      | /     |

## 5.4 Comparison between MEMOTS and 2PPLS on biobjective instances

We realize in this section a comparison between MEMOTS and 2PPLS, for different running times. We do not show the results of the comparison of 2PPLS with other algorithms, since in [59], we already show that MEMOTS gives better values than MOGLS [42] and PMA [46] for the different indicators used.

For MEMOTS, we use the parameters recommended in [59]. The running time of MEMOTS is controlled with the number of iterations, varying between 2000 and 40000. For 2PPLS, the running time is controlled by both $L$ and $N$. We vary $L$ from 4 to 20, and $N$ linearly evolves according to $L$ in the following way:

$$N = 100 + \frac{75}{4}(L - 4) \qquad (3)$$

(in this way, for $L=4$, N=100 and for $L=20$, N=400).

The results are presented in Figure 4 where the evolutions of $D_1$ and $P_{Y_N}$ according to the running time are showed.

We see that except with small running times, the results obtained with 2PPLS are better than with MEMOTS. With 2PPLS, we can generate, for the 250-2 instance, about 90% of the non-dominated points, for the 500-2 instance, about 70% and for the 750-2 instance, about 20%, in reasonable running times. The running times are remarkable since, Mavrotas *et al.* [63] could generate for the 250-2 instance, 81% of $X_E$, but in about 30 minutes, while we can attain this result in about 15s. Also, they need 21 hours to generate 93% of $X_E$, while we only need 42 seconds, that is a ratio equal to 1800!

## 5.5 Three-objective instances

We present the results of MEMOTS and 2PPLS in Table 2 for the 250-3 instance. The results of MEMOTS have been obtained with the parameters recommended in [59]. For 2PPLS, we have used the following parameters: $L = 12$ and $N = 200$. The results of 2PPLS correspond to only one run (for computational overhead reason).

We see that the results obtained with 2PPLS are of better quality for all the indicators considered. The number of potentially efficient solutions obtained with 2PPLS (68540 potentially efficient solutions generated!) is more than five times more important than with MEMOTS. But the running time of 2PPLS is very high: about 8 hours! Indeed, the PLS method is stopped only when it is no more possible to find a new non-dominated neighbor from one of the potentially efficient solutions. In case of three-objective instances, there are so many potentially efficient solutions, that the stop condition is only met after a long running time.

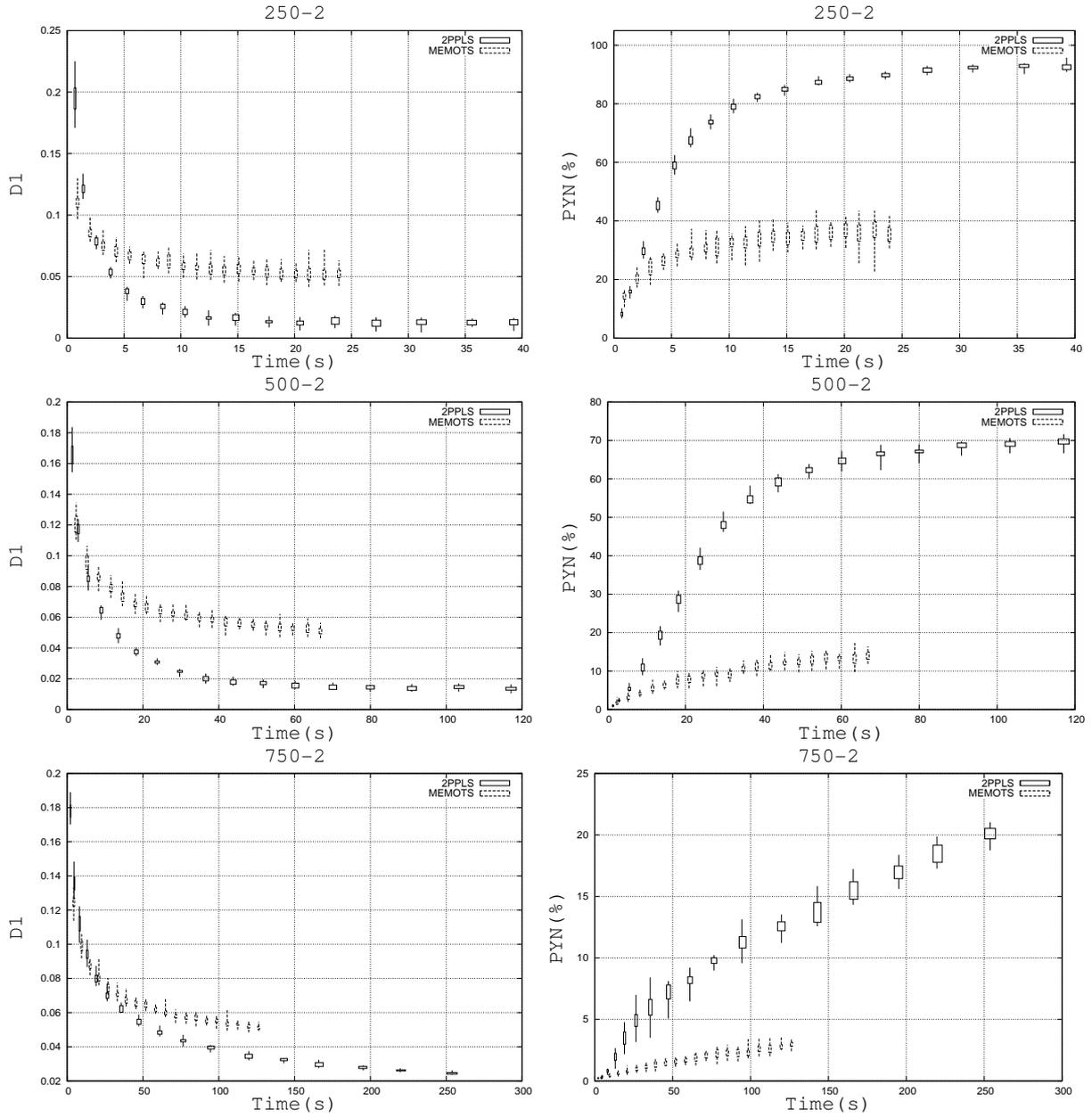

Figure 4: Comparison of MEMOTS and 2PPLS: evolution of $D_1$ and $P_{Y_N}$ according to the running time.

Table 2: Comparison between 2PPLS and MEMOTS for the 250-3 instance (1).

| Instance | Algorithm | $I_{\epsilon 1}$ | $R$ | $D_1$ | $D_2$ | $|PE|$ | Time(s) |
|---|---|---|---|---|---|---|---|
| 250-3 | 2PPLS | **1.017902** | **306.469352** | **3.863** | **4.798** | 68540.00 | 28626.80 |
| | MEMOTS | 1.020264 | 309.680656 | 4.074 | 5.214 | 12043.00 | **129.46** |

From these results, we can say that it will be impossible to use 2PPLS to solve in a reasonable time the 500-3 and 750-3 instances, except if the method is stopped before convergence or if the neighborhood is drastically limited.

We have done that for the 250-3 instance, where we limit the running time of 2PPLS to the same running time than MEMOTS. The results are shown in Table 3, for two different running times.

Table 3: Comparison between 2PPLS and MEMOTS for the 250-3 instance (2).

| Instance | Algorithm | $I_{\epsilon 1}$ | $R$ | $D_1$ | $D_2$ | $|PE|$ | Time(s) |
|---|---|---|---|---|---|---|---|
| 250-3 | MEMOTS | **1.025566** | **313.875552** | **4.283** | **5.989** | 4581.50 | 8.01 |
| | 2PPLS | 1.029983 | 317.772372 | 4.363 | 6.932 | 4872.20 | 8.25 |
| 250-3 | MEMOTS | **1.020264** | 309.680656 | 4.074 | **5.214** | 12043.00 | 129.46 |
| | 2PPLS | 1.020712 | 311.238745 | **4.034** | 5.510 | 13504.35 | 129.18 |

We see that the results obtained with MEMOTS are better than the results of 2PPLS, except when the running time is higher, but only for the $D_1$ indicator. Stopping 2PPLS before convergence can indeed yield results of second-rate quality, since in PLS the exploration of the decision space is not managed as in MEMOTS where the search is constantly guided in regions of low density (in objective space).

We can thus conclude than 2PPLS is more suited to solve biobjective instances.

## 5.6 MOKP

We have also performed some tests for the MOKP ($m = 1$). In this case, we compare our results with the FPTAS of Bazgan *et al.* [10], with $\epsilon = 0.1$, which means that they guarantee a (1, 1)-approximation (the final approximation obtained is however of better quality). We did that for the four types of instances defined by Bazgan *et al.*. The results are provided in Table 5.6.

We see than we obtain better values for $\epsilon$ in less running times (the computer used by Bazgan *et al.* is a 3.4GHz computer with 3072Mb of RAM). Therefore, we can conclude that comparing to a heuristic, the guarantee of performance given by an approximation algorithm has a price.

# 6 Conclusion and perspectives

In this paper, we have first surveyed the vast literature about the MOMKP. We have then proposed 2PPLS with a VLSN to solve the MOMKP. On biobjective instances, 2PPLS with the VLSN that uses a simplified version of MEMOTS to solve the residual problems gives better results than other methods existing in the literature, in particular MEMOTS. Methods based on VLSN for solving MOCO problems are thus a promising stream of research. On the other hand, on three-objective instances, the convergence time of 2PPLS is very high and MEMOTS turns out to be more efficient.

Table 4: Comparison between **2PPLS** and a **FPTAS** for **MOKP** instances.

| Instance | | FPATS | | 2PPLS | |
|---|---|---|---|---|---|
| Type | $n$ | Time(s) | $\epsilon$ | Time(s) | $\epsilon$ |
| A | 400 | 6.084 | 0.0076 | **3.55** | **0.00030** |
| | 700 | 32.275 | 0.0060 | **13.60** | **0.00018** |
| B | 2000 | 1.452 | 0.0028 | **1.292** | **0.00055** |
| | 4000 | 11.220 | 0.0023 | **9.76** | **0.00026** |
| C | 300 | 10.099 | 0.0096 | **9.18** | **0.0011** |
| | 500 | 44.368 | 0.0064 | **23.09** | **0.00076** |
| | 100 | 2.356 | 0.0183 | **0.8379** | **0.0081** |
| D | 200 | 36.226 | 0.0117 | **9.9198** | **0.0027** |
| | 250 | 62.970 | 0.0098 | **18.2884** | **0.0026** |

We think that in the case of randomly generated instances of the MOMKP with more than two objectives, the decision maker should get involved in the process of generating solutions, at the beginning or during the execution of the algorithm, in order to direct the search and to limit the number of solutions generated. An a posteriori approach should be avoided as often as possible. Indeed, if 2PPLS revealed very competitive on biobjective instances, on three-objective instances, as the search in 2PPLS is not as well-directed than in MEMOTS or than in the other memetic algorithms using scalarizing functions to guide the search, the results with MEMOTS were better. That corroborates the need of the intervention of the decision maker during the process of 2PPLS and to thus modify this approach in an interactive way.

## Acknowledgments

T. Lust thanks the "Fonds National de la Recherche Scientifique" for a research fellow grant (Aspirant FNRS). We also thank V. Barichard, D.S. Vianna, M. João Alves, R. Beausoleil and A. Alsheddy for having provided us their results.